\documentclass[12pt]{iopart}

\usepackage{graphicx}
\usepackage{color}

\usepackage{iopams}

\usepackage{amscd}

\makeatletter

\@addtoreset{equation}{section}
 
\makeatother

\newcommand{\beq}{\begin{equation}}
\newcommand{\eeq}{\end{equation}}
\newcommand{\beqa}{\begin{eqnarray}}
\newcommand{\eeqa}{\end{eqnarray}}
\newcommand{\noi}{\noindent}

\newcommand{\g}{{\mathfrak g}}

\def\>{\rangle}
\def\<{\langle}

\begin{document}

\title[]{ Unexpected Features of Supersymmetry with   Central Charges  }

\author{Rutwig Campoamor-Stursberg\dag}
\address{\dag\ I.M.I-U.C.M,
Plaza de Ciencias 3, E-28040 Madrid, Spain}
\ead{rutwig@pdi.ucm.es}

\author{Michel Rausch de Traubenberg\ddag}
\address{\ddag IPHC-DRS, UdS, CNRS, IN2P3, 23  rue du Loess,
F-67037 Strasbourg Cedex, France}
\ead{Michel.Rausch@IReS.in2p3.fr}

\begin{abstract}
It is shown that $N=2$ supersymmetric theories with central
charges present some hidden quartic symmetry. This enables us to
construct representations of the quartic structure induced by
superalgebra representations.
\end{abstract}


\maketitle

\section{Motivation and introduction}

For a long time, Lie algebras and superalgebras have been the main
group-theoretical tool for the description of physical phenomena,
leading to the belief that binary structures are the only adequate
algebraic tool capable of reflecting a realistic description of
processes in physics. However, evolution of the theoretical
background and more recent formal developments and, in particular,
efforts towards a unified and effective description of
interactions has shown that it is convenient, if not unavoidable,
to consider algebras beyond the binary structures. Examples of
this generalized algebraic approach are given {\it e.g.} by the
$n$-linear algebras in Quantum Mechanics \cite{n}, ternary
algebras in the description of multiple $M_2-$branes \cite{gbl,az}
or higher order extensions of the Poincar\'e algebra \cite{flie}.

Among these possibilities, the $F$-ary extensions ($F>2$) of Lie
superalgebras (called Lie algebras of order $F$) introduced and
analysed in \cite{flie,gr,hopf,color}, offered the possibility of
defining higher order extensions of the Poincar\'e algebra based
on solid physical arguments. In particular, a specific cubic
extension in arbitrary space-time dimensions was shown to be of
interest in the frame of Quantum Field Theory
\cite{cubic1,cubic2,p-form}. In addition, it was possible to
conceive and construct the notions of  group \cite{hopf} and
adapted superspace associated to these structures \cite{super}.

In this work we show that certain graded Lie superalgebras might
induce specific quartic extensions of Lie algebras. This
construction is then applied to show that we can associate
naturally a quartic extension of the Poincar\'e algebra to the
standard  $N=2$ supersymmetric extensions (with central charges).
In this construction, the central charges turn out to play a key
role, as they constitute the essential  ingredient to introduce
the notion of a hidden quartic symmetry. It follows that massive
invariant $N=2$ Lagrangians are also invariant with respect to
these hidden symmetries. We finally conclude that the quartic
extensions obtained by this method give rise to a hierarchy of
representations that emerges from the standard representation of
the corresponding supersymmetric theory.

\section{Quartic  structures induced by quadratic algebras} \label{sec}

The $F-$order structures were introduced mainly with the purpose
of providing new non-trivial higher order (order greater that two)
extensions of the Poincar\'e algebra  involving generators $Q$
which can be seen as ``an $F-$root'' of the space-time
translations (see {\it e.g.} (\ref{4poin}) or (\ref{D=9-10}) when
$F=4$) \cite{flie}. In this work we are mainly interested in the
quartic case. Consequently, we briefly recall the salient
properties of these structures.

\subsection{Quartic extensions of the Poincar\'e algebra}\label{sec:quart}
The vector space $\g=\g_0 \oplus \g_1 = \langle X_i, i=1,\cdots,
\dim \g_0\rangle \oplus\langle Y_a, a=1,\cdots, \dim \g_1\rangle
$\footnote{We would like to point out that the symbol $\oplus$
here and in the sequel can be understood as a direct sum of vector
spaces. } is called an elementary Lie algebra of order four if it
satisfies the following brackets

\beqa \label{eq:4-alg} \big[X_i,X_j\big]=f_{ij}{}^k X_k, & & \ \
\big[X_i,Y_a\big]=R_{ia}{}^b Y_b, \ \ \nonumber \\
\big\{Y_{a_i},Y_{a_2},Y_{a_3},Y_{a_4}\big\}&=& \sum
\limits_{\sigma \in S_4}
Y_{\sigma(a_1)}Y_{\sigma(a_2)}Y_{\sigma(a_3)}Y_{\sigma(a_4)} =
Q_{a_1 a_2 a_3 a_4}{}^i X_i, \eeqa

\noi $S_4$ being the permutation group with four elements. In
addition, we have also the following generalised Jacobi
identities:

\beqa \label{eq:J} \fl \big[Y_{a_1},
\big\{Y_{a_2},Y_{a_3},Y_{a_4},Y_{a_5} \big\}\big]+ \big[Y_{a_2},
\big\{Y_{a_3},Y_{a_4},Y_{a_5},Y_{a_1} \big\}\big]+
\big[Y_{a_3}, \big\{Y_{a_4},Y_{a_5},Y_{a_1},Y_{a_2} \big\}\big]+\nonumber \\
\fl \big[Y_{a_4}, \big\{Y_{a_5},Y_{a_1},Y_{a_2},Y_{a_3}
\big\}\big] + \big[Y_{a_5}, \big\{Y_{a_1},Y_{a_2},Y_{a_3},Y_{a_4}
\big\}\big] =0. \eeqa

Along the lines of the algebraic structure (\ref{eq:4-alg}), there
is the possibility of constructing quartic extensions of the
Poincar\'e algebra in arbitrary space-time dimensions. Within this
frame, two quartic extensions of the Poincar\'e algebra will be
considered in detail: that in $D=4$ for obvious physical reasons,
as well as $D=10$, which will be shown later (Section
\ref{sec:4susy}) to constitute an exceptional case.

The quartic extensions of the Poincar\'e algebra in $D=4$
dimensions are constructed by considering two Majorana
spinors\footnote{For a brief summary of the properties of spinors,
see section \ref{sec:spin}.}. In the $\mathfrak{sl}(2,\mathbb C)
\cong \mathfrak{so}(1,3)$ notations of dotted and undotted
indices, a left-handed spinor is given by $\psi_L{}^\alpha$ and  a
right-handed spinor by $\bar \psi_R{}_{\dot \alpha}$.  The spinor
conventions to raise/lower indices are as follows $\psi_L{}_\alpha
=\varepsilon_{\alpha\beta}\psi_L{}^\beta$, $\psi_L{}^\alpha
=\varepsilon^{\alpha\beta}\psi_L{}_\beta$,
$\bar\psi_R{}_{\dot\alpha}=\varepsilon_{\dot\alpha
\dot\beta}\bar\psi_R{}^{\dot\beta}$, $\bar\psi_R{}^{\dot\alpha}
=\varepsilon^{\dot\alpha\dot\beta}\bar\psi_R{}_{\dot\beta}$ with
$(\psi_\alpha)^\star =\bar\psi_{\dot\alpha}$, $\varepsilon_{12} =
\varepsilon_{\dot 1\dot 2}=1$, $\varepsilon^{12} =
\varepsilon^{\dot 1\dot 2}=-1$. The $4D$ Dirac matrices,  in the
Weyl representation,  are
\begin{equation}
\label{eq:gamma} \Gamma^\mu = \left(
\begin{array}{cc}
 0 & \sigma^\mu\\
 \bar\sigma^\mu & 0
\end{array}\right),
\end{equation}
with $\sigma^\mu{}_{\alpha\dot\alpha}=(1,\sigma^i ),
 \; \bar\sigma^{\mu\dot\alpha\alpha}=
 (1,-\sigma^i)$,
$\sigma^i$ ($i=1,2,3$) being the Pauli matrices. With these
notations, we introduce two series of Majorana spinors
$Q^I{}_{\alpha}, \bar Q_I{}_{\dot \alpha}$  satisfying the
relation $(Q^I{}_{\alpha})^\dag= \bar Q_I{}_{\dot \alpha}$. The
Lie algebra of order four with $\g_0=I{\mathfrak{so}}(1,3) $ (the
Poincar\'e algebra) and $\g_1= \big<Q^I{}_\alpha, \bar Q_{I \dot
\alpha}\big>$ define the following quartic extension of the
Poincar\'e algebra\footnote{For shortness, we only give the
quartic brackets explicitly.}

\beqa \label{4poin} \fl
\left\{Q^{I_1}{}_{\alpha_1},Q^{I_2}{}_{\alpha_2},
Q^{I_3}{}_{\alpha_3}, Q^{I_4}{}_{\alpha_4}
\right\}&=&0, \nonumber \\
\fl \left\{Q^{I_1}{}_{\alpha_1},Q^{I_2}{}_{\alpha_2},
Q^{I_3}{}_{\alpha_3}, \bar{Q}_{I_4}{}_{\dot \alpha_4} \right\}&=&
2i\Big(\delta^{I_1}{}_{I_4} \varepsilon^{I_2 I_3}
\varepsilon_{\alpha_2 \alpha_3} \sigma^\mu{}_{\alpha_1 \dot
\alpha_4} +\delta^{I_2}{}_{I_4} \varepsilon^{I_1 I_3}
\varepsilon_{\alpha_1 \alpha_3} \sigma^\mu{}_{\alpha_2 \dot \alpha_4}  \\
&+&\delta^{I_3}{}_{I_4} \varepsilon^{I_1 I_2}
\varepsilon_{\alpha_1 \alpha_2} \sigma^\mu{}_{\alpha_3 \dot
\alpha_4}\Big) P_\mu,
\nonumber \\
\fl \left\{Q^{I_1}{}_{\alpha_1},Q^{I_2}{}_{\alpha_2},
\bar{Q}_{I_3}{}_{\dot \alpha_3}, \bar{Q}_{I_4}{}_{\dot \alpha_4}
\right\}&=& 0, \nonumber \eeqa

\noi the remaining brackets involving three $\bar Q$ and one $Q$
or four $\bar Q$ being obtained immediately (the tensor
$\varepsilon^{IJ}$ is defined by
$-\varepsilon^{12}=\varepsilon^{21}=\varepsilon_{12}=-\varepsilon_{21}=1$).

\smallskip

The quartic extension of the Poincar\'e algebra in  $D=10$ is
constructed by considering a Majorana spinor    $Q_A$
($\g_0=I{\mathfrak{so}}(1,9)$ and $\g_1=\big<Q_A\big>)$. We denote
$A=(a,a')$ the spinor indices, with $a$ (resp. $a'$) representing
the indices for the left- (resp. right-)handed part of the spinor
$Q_A$. With these conventions, we introduce $Q^+_a$ the
left-handed and $Q^-_{a'}$ the right handed part of $Q$, so that
the quartic part of the algebra takes the form

\beqa \label{D=9-10}
\fl \{Q^+_{a_1}, Q^+_{a_2},Q^+_{a_3},Q^+_{a_4}\}&=&0, \nonumber\\
\fl \{Q^+_{a_1}, Q^+_{a_2},Q^+_{a_3},Q^-_{ a'_4}\}&=&2i
\Big(C^{+-}_{a_1 a'_4} ( \Sigma^\mu C^{-+})_{a_2 a_3}  +
C^{+-}_{a_2 a'_4} ( \Sigma^\mu C^{-+})_{a_1 a_3}
+ C^{+-}_{a_3 a'_4} (\Sigma^\mu C^{-+})_{a_1 a_2}\big) P_\mu, \nonumber \\
\fl \{Q^+_{a_1}, Q^+_{ a_2},Q^-_{a'_3}, Q^-_{a'_4}\}&=&0,
 \eeqa
with \beqa \Gamma^\mu = \left(\begin{array}{cc}
0&\Sigma^\mu\\\tilde
\Sigma^\mu&0 \end{array}\right), \ \ C=\left(\begin{array}{cc} 0&C^{+-} \\
C^{-+}&0\end{array}\right), \nonumber \eeqa being the
$10-$dimensional Dirac and the charge conjugation matrices
respectively.

\noi It has to be emphasized that the algebraic structure defined
in this manner is neither an algebra nor a $4-$algebra in the
usual sense, but a kind of hybrid structure. Indeed, some of the
brackets are quadratic $[\g_0, \g_0] \subseteq \g_0, [\g_0, \g_1]
\subseteq \g_1$, while some others are quartic
$\big\{\g_1,\g_1,\g_1,\g_1 \big\} \subseteq \g_0$. This feature
represents one of the difficulties to handle with these algebraic
structures. \noi Our aim is to analyse to which extent the quartic
bracket given in (\ref{eq:4-alg}) can be obtained by means of
appropriate quadratic brackets. It is worthy to be remarked that
various attempts to inspect the same problem for cubic instead of
quartic extensions did not succeed. The obstructions encountered
may provide, in some sense, a structural explanation for the
difficulties found in the various constructions of the cubic
extensions of the Poincar\'e algebra, despite of some interesting
results obtained with such cubic structures.

\subsection{Relationship between quartic algebras and Lie superalgebras}

The study of relationships between quadratic and higher order
algebras is certainly not a new problem. For instance, it has
already been established in \cite{fo} that some ternary algebras
of the Filippov type considered in  the Bagger-Lambert-Gustavsson
model are equivalent to certain Lie (super-)algebras
\cite{gbl,f,fo}.

We will prove that a similar result can be obtained in this case
and that certain types of graded-Lie superalgebras actually induce
a quartic structure in analogy to those given in (\ref{eq:4-alg}),
but with slight formal differences. This will enable us to develop
a procedure to associate quartic algebras to binary algebras.
There is a natural comparison of this construction and the
induction theorem of the second paper of \cite{flie} that allows
to construct higher order Lie algebras starting from arbitrary Lie
(super)algebras. The main difference with respect to this approach
is that in the present ansatz higher order brackets are compatible
with quadratic brackets.\\

The starting point for our construction is a $\mathbb Z_2 \times
\mathbb Z_2-$graded Lie superalgebra \beqa\g= \big(\g_{(0,0)}
\oplus \g_{(1,1)}\big) \oplus\big(\g_{(1,0)} \oplus
\g_{(0,1)}\big), \eeqa where $(a,b) \in \mathbb Z_2 \times \mathbb
Z_2$ and $\g_{(a,b)}$ is even (resp. odd) when $a+b=0
\;\rm{mod}\;2$ (resp. $a+b=1 \;\rm{mod} \; 2$). In the following
$\g_0=\g_{(0,0)}\oplus \g_{(1,1)}$ (resp. $\g_1=\g_{(0,1)}\oplus
\g_{(1,0)}$) will always denote the even (resp. odd) sector of the
algebra. Considering the corresponding bases for the grading
blocks: \beqa \label{eq:gradedLie}
\begin{array}{ll}
\g_{(0,0)}=\langle B_i, i=1,\cdots,\dim \g_{(0,0)}\rangle,&
\g_{(1,1)}=\langle Z\rangle, \\
\g_{(1,0)}=\langle F^+_a, a=1,\cdots,\dim \g_{(1,0)}\rangle,&
\g_{(0,1)}=\langle F^-_a, a=1,\cdots,\dim \g_{(0,1)}\rangle,
\end{array}
\eeqa  the corresponding commutation relations are \beqa
\label{eq:gradedsuper}
\begin{array}{ll}
\big[B_i,B_j\big]=f_{ij}{}^k B_k,&[B_i,Z]=0,\\
\big[B_i,F^{\varepsilon}_a]=R^{\varepsilon}_i{}_a{}^b F^{\varepsilon}_b,& \big[Z,F^{\varepsilon}_a]=0,\\
\big\{F^{\varepsilon}_i,F^{\varepsilon}_j\big\}=Q^{\varepsilon}_{ij}{}^a
B_a,& \big\{F^{+}_i,F^{-}_j\big\}=g_{ij} Z,
\end{array}
\eeqa  where $\varepsilon=\pm$. Furthermore, the superalgebra
defined by (\ref{eq:gradedsuper}) satisfies also the appropriate
Jacobi identities (that we do not recall here). It is important to
notice that $\g_{(1,1)}$ commutes with all remaining factors, in
other words, that $Z$ acts like a central charge. Formally, there
is no difficulty to generalise this structure to have a higher
number of central charges $Z_i,i=1,\cdots,k$, so that
$\g_{(1,1)}=\left<Z_1,\cdots,Z_k\right>$. In this work, however,
we focus on the applications of the construction to the case of
$N=2$ supersymmetric extensions of the Poincar\'e algebra, so that
only one central charge is required. In the following, we only
consider the case where $\g_{(1,1)}$ is of dimension one. On the
other hand, we could have considered the following grading blocks:
$\g_{(0,0)}=\langle Z\rangle, \g_{(1,1)}=\langle B_i,
i=1,\cdots,\dim \g_{(1,1)}\rangle$. In this case, the fermionic
part of the brackets reads \beqa \label{eq:gradedsuper2}
\big\{F^{+}_i,F^{-}_j\big\}=Q_{ij}{}^a B_a,&
\big\{F^{\varepsilon}_i,F^{\varepsilon}_j\big\}=g^\varepsilon_{ij}
Z, \ \ \ \varepsilon = \pm. \eeqa

With these preliminary assumptions, we are in situation of
establishing the main result of this section by explicitly
constructing  a quartic algebra associated to the superalgebra
(\ref{eq:gradedLie}). This structure, in the light of equation
(\ref{eq:4-alg}), enables us to  express the non-graded part of
the algebra in terms of the graded part. In fact, using the
obvious relation, \beqa \fl
\big\{A_1,A_2,A_3,A_4\big\}=\big\{\big\{A_1,A_2\big\},\big\{A_3,A_4\big\}
\big\} + \big\{\big\{A_1,A_3\big\},\big\{A_2,A_4\big\} \big\}+
\big\{\big\{A_1,A_4\big\},\big\{A_2,A_3\big\} \big\}, \nonumber
\eeqa
 the relations
\beqa \label{eq:quart} \fl
\big\{F^+_{a_1},F^+_{a_2},F^+_{a_3},F^+_{a_4}\big\} &=&
\big(Q^+_{a_1 a_2}{}^i Q^+_{a_3 a_4}{}^j + Q^+_{a_1 a_3}{}^i
Q^+_{a_4 a_2}{}^j + Q^+_{a_1 a_4}{}^i Q^+_{a_2 a_3}{}^j \big)
\big\{B_i,B_j\big\} \nonumber \\
\fl \big\{F^+_{a_1},F^+_{a_2},F^+_{a_3},F^-_{a_4}\big\} &=&
2Z(g_{a_1 a_4} Q^+_{a_2 a_3}{}^i + g_{a_2 a_4} Q^+_{a_1 a_3}{}^i +
g_{a_3 a_4} Q^+_{a_1 a_2}{}^i ) B_i  \\
\fl \big\{F^+_{a_1},F^+_{a_2},F^-_{a_3},F^-_{a_4}\big\}&=&
Q^+_{a_1 a_2}{}^i Q^-_{a_3 a_4}{}^j \big\{B_i,B_j\big\} +
2\big(g_{i_1 i_3} g_{i_2i_4} +g_{i_1 i_4} g_{i_2i_3}\big)Z^2   \
,\nonumber \eeqa (plus similar relations involving either three
$F^-$ and one $F^+$ or four $F^-$) follow at once. Had we chosen
the alternative possibility for the fermionic brackets given
above, the brackets (\ref{eq:quart}) would be subjected to a
corresponding minor modification.

\medskip

Some comments are in order here. The superalgebra reproduces the
algebra of type (\ref{eq:4-alg}), but  in a slightly modified
form. Since we are constructing an analogue of the four-Lie
algebra (\ref{eq:4-alg}), we also assume that the algebra
associated to the Lie superalgebra (\ref{eq:gradedsuper}) inherits
the same algebraic structure. More precisely, we suppose that the
algebra is partially quadratic and partially quartic, that is,
$[\g_0,\g_0] \subseteq \g_0, [\g_0,\g_1] \subseteq \g_1$ (these
brackets are the same of the corresponding brackets of the Lie
superalgebra), but now the quartic brackets
$\{\g_1,\g_1,\g_1,\g_1\}$ close quadratically in $\g_0$. This
means that the structure which emerges in this process closes in
the universal enveloping algebra of $\g_0$ since the R.H.S.
involves symmetric products of elements of $\g_0$. For this reason
it could be called a non-linear (or quadratic) Lie algebra of
order four. To round off the relationship between Lie
superalgebras and the algebra defined by the relations
(\ref{eq:quart}), we have to check that the Jacobi identities of
Lie superalgebras reproduce the generalised Jacobi identity
(\ref{eq:J}).  We observe that there is no need for this to hold
in full generality, as there is no reason for the relations
(\ref{eq:gradedsuper}), together with the Jacobi identities of Lie
superalgebras, to imply the identity (\ref{eq:J}). However, it
happens that if we have a finite dimensional representation of
(\ref{eq:quart}), the identities (\ref{eq:J}) are trivially
satisfied. For the case under inspection in this work this will
not be a constraint, since the generalised Jacobi identity will be
trivially satisfied as well. This happens because the
four-brackets $\{\g_1,\g_1,\g_1,\g_1\}$ close upon $P_\mu$ or $Z$
(see below) thus we automatically have
$[\{\g_1,\g_1,\g_1,\g_1\},\g_1]=0$.
 Finally,
since the quadratic relations (\ref{eq:gradedsuper}) imply the
quartic relations (\ref{eq:quart},  the superalgebra is
compatible with the quartic algebra structure described by
equations (\ref{eq:quart}), meaning that any algebra of the form
(\ref{eq:gradedLie}) satisfying the (anti)commutation  relations
(\ref{eq:gradedsuper}) {\it automatically satisfies}, by
construction, the relations (\ref{eq:quart}).

\medskip

This last observation has an interesting consequence. It is well
known that Lie (super)algebras correspond to infinitesimal
transformations and that one is able to associate finite
dimensional transformations having the structure of groups and
leading to the notion of Lie (super-)groups. It has been further
proved that  using  heavy algebraic machinery \cite{hopf} there is
a way to associate an appropriate group to higher order Lie
algebras. The benefit of the previous construction is that
 one can associate a group to the quartic algebra using standard procedures,
and therefore avoiding complicate formal tools. This means, in
particular, that this  direct approach reproduces  the standard
quantum mechanical formalism for symmetry descriptions as
presented {\it e.g} in \cite{w} (pp. 50-55). On the other hand, it
constitutes a well established fact that faithful representations
of higher order algebras are infinite dimensional \cite{flie,
hopf}, which immediately implies that matrix representations are
automatically non-faithful. The compatibility of quadratic
relations with quartic relations further  means that any
representation of the Lie superalgebra (\ref{eq:gradedLie}) will
also be a (non-faithful) representation of the quartic algebra
(\ref{eq:quart})\footnote{\label{foot} If we denote $\g_{(2)}$
(resp. $\g_{(4)}$) the superalgebra (\ref{eq:gradedLie}) (resp. of
the quartic algebra (\ref{eq:quart})) and  define their associated
universal enveloping algebra ${\cal U}(\g_{(2)})$ (resp. ${\cal
U}(\g_{(4)}))$ $-$for the definition of the enveloping algebra of
higher order algebras see \cite{flie,hopf}$-$), the
representations of $\g_{(2)}$ (resp. $\g_{(4)}$) extend to
representations of ${\cal U}(\g_{(2)})$ (resp. ${\cal
U}(\g_{(4)}))$. As a consequence of the quotient ${\cal
U}(\g_{(4)})/{\cal I}_{(2)} \cong {\cal U}(\g_{(2)})$ (where
${\cal I}_{(2)}$ is the two-sided ideal generated by the relations
(\ref{eq:gradedsuper}) - or (\ref{eq:gradedsuper2}) -) it turns
out that a representation of $\g_{(2)}$ is also a representation
of $\g_{(4)}$. }. As expected, the converse is not necessarily
true.\newline An unsuspected consequence of the construction above
is that it suggests the existence of some hidden quartic
symmetries in superalgebras\footnote{Successive attempts to adapt
this methodology to the framework of colored Lie superalgebras
\cite{col}, have led to serious formal obstructions that cannot be
surmounted.}. In general terms, this construction can be
understood, in some sense,  as a ``square'' of the
graded-superalgebra (\ref{eq:gradedsuper}). By this we
specifically mean that we can naturally associate to the graded
superalgebra (\ref{eq:gradedsuper}) the quartic algebra
(\ref{eq:quart}). It is in this sense that the term ``hidden
quartic symmetry" appears in the usual framework of graded Lie
superalgebras.

\section{Supersymmetry and quartic extensions of the Poincar\'e algebra}\label{sec:4susy}
Bearing in mind that we are mainly interested on (extended)
space-time symmetries compatible with the principle  of relativity
and quantum mechanics, it is worthy to be inspected in detail
whether higher order symmetries emerge in supersymmetry. In this
section, we first recall the properties of spinor in arbitrary
space-time dimensions. In the next subsection, we briefly recall
the principle for the construction of supersymmetric theory in
arbitrary space-time dimensions and address the question whether
these algebraic structures are of the form  (\ref{eq:gradedsuper})
or (\ref{eq:gradedsuper2}). We finally show to which extent
quadratic extensions of the Poincar\'e algebra can be naturally
associated to the usual supersymmetric extensions.

\subsection{Properties of spinors in arbitrary space-time dimensions}
\label{sec:spin} For our purpose it is convenient to review the
main properties of spinors and Dirac-$\Gamma$ matrices in any
dimensions. Introducing the tensor metric in $D-$space-time
dimensions $\eta_{\mu \nu}=\rm{diag}(1,-1,\cdots,-1)$, the Dirac
$\Gamma-$matrices are defined by, \beqa
\left\{\Gamma_\mu,\Gamma_\nu\right\}=2 \eta_{\mu \nu}.\nonumber
\eeqa One can easily show  that the Dirac matrices are complex
$2^{[D/2]}\times 2^{[D/2]}$ matrices ($[a]$ representing the
integer part of $a$). These matrices act on Dirac spinors
$\Psi_D$. Such  spinors are complex and exist in any space-time
dimensions. Furthermore, when the dimension is even, on can define
the chirality matrix, $\chi=i^{[D/2]}\Gamma_0 \cdots \Gamma_{D-1}$
($\chi^2=1, \{\Gamma_\mu,\chi\}=0)$ leading to the so-called Weyl
spinors $\Psi_\pm=1/2(1\pm\chi)\Psi_D$. In certain dimensions,
additional specific spinors can be defined. For that purpose, we
introduce a matrix $B$ such that, \beqa B\Gamma_\mu B^{-1} = \pm
\Gamma^\mu{}^\star, \noindent \eeqa where $\Gamma^\star$ denotes
the complex conjugate of the matrix $\Gamma$. The sign in the
equation above depends on the space-time dimension. When the
matrix $B$ satisfies the property  $BB^\star=1$, a Majorana spinor
can be defined. A Majorana spinor  exists in certain dimensions
and satisfies the condition $\Psi^\star = B \Psi$. In such a case
one can find a realisation of the Dirac matrices where all the
matrices are real (or purely imaginary), and as a consequence the
components of a Majorana spinor can be chosen to be real. On the
contrary, when $B B^\star=-1$, an $SU(2)-$Majorana spinor $\Psi^i,
i=1,2$ can be defined. An $SU(2)-$Majorana spinor
 satisfies $\Psi_i^\star = \epsilon_{ij} B\Psi^j$
with $\epsilon_{ij}$ the $SU(2)$ invariant volume form. The
various types of spinors strongly depend on the space-time
dimension. It is however important to emphasize that the results
are indeed periodic of period eight. As simple consequence, a
property valid in $D-$space-time dimensions is also valid in
$(D+8)-$dimensions. For instance, a Majorana spinor in $D=3$ can
be defined, but also in $D=11$. Table \ref{fig:spin}  summarizes
the type of spinors, given modulo $8$, that is, from $D=0$ mod $8$
to $D=7$ mod $8$.
\bigskip

\begin{center}
\begin{table}[!h]
\begin{indent}
\caption{Types of spinors in various dimensions, $N_f$ indicates
the number of real components of a given spinor.}
{\tiny\label{fig:spin} \medskip
\begin{tabular}{|c|c|c|c|c|c|c|c|}
\hline $D=0\; \rm{mod} \;8$&$D=1\; \rm{mod} \;8$&$D=2\; \rm{mod}
\;8 $&$D=3\; \rm{mod} \;8$
&$D=4\; \rm{mod} \;8$&$D=5\; \rm{mod} \;8$&$D=6\; \rm{mod} \;8$&$D=7\; \rm{mod} \;8$\\
\hline Majorana&Majorana&\begin{tabular}{c}Majorana-\\Weyl
\end{tabular}&Majorana&Majorana&
\begin{tabular}{c}$SU(2)$-\\Majorana\end{tabular}&
\begin{tabular}{c}$SU(2)$-\\Majorana-\\Weyl\end{tabular}&
\begin{tabular}{c}$SU(2)$-\\Majorana\end{tabular}
\\
\hline $N_f= 2^{\frac{d}{2}}$&$N_f=2^{\frac{d-1}{2}}$&$N_f=\frac12
2^{\frac{d}{2}}$&
$N_f=2^{\frac{d-1}{2}}$&$N_f=2^{\frac{d}{2}}$&$N_f=2^{\frac{d-1}{2}}$&
$N_f=\frac12 2^{\frac{d-1}{2}}$&$N_f=2^{\frac{d}{2}}$
\\ \hline
\end{tabular}
}
\end{indent}
\end{table}
\end{center}

\noindent If we introduce $\Gamma^\mu$ the Dirac $\Gamma-$matrices
in $D$ space-time dimensions and $C$ the charge conjugation matrix
defined  by \beqa C \Gamma^\mu C^{-1} = \pm \Gamma^{\mu t}, \eeqa
where $\Gamma^t$ denotes the transpose of the matrix $\Gamma$ and
the sign depends on the space-time dimension, the matrices
$\Gamma^\mu C$ and $C$  are either symmetric or anti-symmetric
depending on the dimension as indicated  in Table \ref{fig:sym}
(the results are also periodic of period eight).

\begin{center}
\hskip-.5truecm
\begin{table}[!h]
\caption{Symmetry of the $\Gamma-$matrices}\label{fig:sym}
{\footnotesize
\begin{tabular}{|c|c|c|c|c|c|c|c|c|}
\hline
$D \; \rm{mod} \; 8$&$0$&$1$&$2$&$3$&$4$&$5$&$6$&$7$\\
\hline
$C$&sym&sym&\begin{tabular}{c}sym\\anti-sym\end{tabular}&anti-sym&anti-sym&
anti-sym&\begin{tabular}{c}sym \\anti-sym\end{tabular}&sym\\
\hline $\Gamma^\mu C$&\begin{tabular}{c}sym
\\anti-sym\end{tabular} &sym&sym&sym&\begin{tabular}{c}sym
\\anti-sym\end{tabular}&
anti-sym&anti-sym&anti-sym\\
\hline
\end{tabular}
}
\end{table}
\end{center}

\noindent Finally, we recall that the tensor product of two Weyl
spinors for $D=2n$ decomposes on the set of $p-$forms ($[p]$
denotes $p-$forms and $[n]_\pm$  (anti-)self-dual $n-$forms)

\beqa \label{tensor2} {\cal S}_+ \otimes  {\cal S}_+ = \left\{
\begin{array}{ll} \big[0\big] \oplus \big[2\big] \oplus \cdots
\oplus \big[n\big]_+ & {\rm  when\; } n \; {\rm is\; even} \cr
\big[1\big] \oplus \big[2\big] \oplus \cdots \oplus \big[n\big]_+&
{\rm  when\; } n \; {\rm is\; odd}
\end{array} \right. ,\nonumber \\
\\
{\cal S}_+ \otimes  {\cal S}_- = \left\{ \begin{array}{ll}
\big[1\big] \oplus \big[3\big]  \oplus\cdots \oplus \big[n-1\big]&
{\rm  when\; } n \; {\rm is\; even} \cr \big[0\big] \oplus
\big[2\big] \oplus\cdots \oplus \big[n-1\big]& {\rm  when\; } n \;
{\rm is\; odd}
\end{array} \right. ,\nonumber
\eeqa with ${\cal S}_\pm$ denoting the left- (and right-)handed
spinors.\newline These results constitute standard material and
can be found in many textbooks on the subject (see {\it e.g.}
\cite{cliff}).

\subsection{Supersymmetry in any space-time dimensions: a brief summary}
Adjoining to the Poincar\'e generators   spinors of the form given
in Table \ref{fig:spin}, one is able to construct a supersymmetric
extension of the Poincar\'e algebra in space-time dimensions $D\le
11$ (see {\it e.g.} \cite{s} for a systematic study of
supersymmetry in arbitrary space-time dimensions). In order to
obtain a quartic extension of the Poincar\'e algebra associated to
supersymmetry along the lines of Section \ref{sec},  we have to
check whether or not the various superalgebras are of the form
(\ref{eq:gradedsuper}) or (\ref{eq:gradedsuper2}). This means in
particular that since $\g_{(1,1)}=\left<Z\right>\ne \emptyset$ -
or $\g_{(0,0)}=\left<Z\right>\ne \emptyset$ - at least one central
charge is needed, and thus  at least an $N=2$ supersymmetry.
Inspecting the $D-$dimensional supersymetric extensions of the
Poincar\'e algebra with the help of Table \ref{fig:sym}  and
playing with the central charges \cite{s}, one observes that the
$N=2$ supersymmetric algebra may be put on the form of
(\ref{eq:gradedLie}) (or its modified version) in all dimensions
but $D=5,11$. It turns out that the $D=8,10$ cases are exceptional
since only one Majorana
 spinor is required. In order to illustrate the procedure we focus on
 the $D=10$ case. In order to reproduce the
superalgebra of the form (\ref{eq:gradedsuper}) we have to
consider type $IIA$ supersymmetry (type $I$ and $IIB$ are
excluded). Type $IIA$ supersymmetry is based on a Majorana spinor
(or equivalently one left-handed and one right-handed
Majorana-Weyl spinors). With the notations of (\ref{D=9-10}) we
have $\g_0=\g_{(0,0)} \oplus \g_{(1,1)}=I{\mathfrak{so}}(1,9)
\oplus\big<Z\big>, \g_1=\g_{(1,0)}\oplus
\g_{(0,1)}=\big<Q_a^+\big>\oplus\big<Q_{a'}^-\big>$ and the
fermionic part of  the algebra takes the form

\beqa \fl \big\{Q^+_a, Q^+_b\big\}= 2i( \Sigma^\mu C^{-+})_{ab}
P_\mu, \ \
 \big\{Q^-_{a'}, Q^-_{b'}\big\}= 2i(\tilde \Sigma^\mu C^{+-})_{a'b'} P_\mu, \ \
\big\{Q^+_a, Q^-_{b'}\big\}=  Z C^{+-}_{ab'}. \eeqa

\subsection{Quartic extension of the Poincar\'e algebra associated to supersymmetry}

After these considerations we are now in situation of presenting
our main result. When the $N=2$ superalgebras with one central
charge of the previous section can be put on the form of
(\ref{eq:gradedsuper}) (or (\ref{eq:gradedsuper2})), as  we have
seen, they enable us to propose an alternative construction of
quartic algebras. Let us stress that the role of the central
charge is essential  for the argumentation. This specifically
means that up to the dimensions $D=5,11$, quartic extensions of
the Poincar\'e algebra are realisable. An amazing feature of these
considerations is that super-Poincar\'e algebra with central
charges exhibits a hidden quartic  symmetry. The quartic brackets
are constructed along the lines of (\ref{eq:quart}) and it has to
be mentioned that since $P_\mu$ and $Z$ commute with the $Q$'s,
the generalised Jacobi identity (\ref{eq:J}) is satisfied
independently of any representations. Moreover, as already
commented, these algebraic structures associated to Lie
superalgebras are hybrid, partially reflecting the structure of an
algebra and partially of a four-algebra.

As an illustration, we give the structure of the algebra
constructed along these lines for the exceptional case $D=10$.
Using the algebraic structure (\ref{D=9-10}) with $\g_0=\g_{(0,0)}
\oplus \g_{(1,1)}=I{\mathfrak{so}}(1,9) \oplus\big<Z\big>,
\g_1=\g_{(1,0)}\oplus
\g_{(0,1)}=\big<Q_a^+\big>\oplus\big<Q_{a'}^-\big>$ and the
construction given in section \ref{sec:quart} one obtains the
following brackets (we only give the quartic brackets) \beqa \fl
\left\{Q_{a_1}^+,Q_{a_2}^+,Q_{a_3}^+,Q_{a_4}^+\right\}&=& -8\Big(
(\Sigma^\mu C^{-+})_{a_1 a_2} (\Sigma^\nu C^{-+})_{a_3 a_4} +\
(\Sigma^\mu C^{-+})_{a_1 a_3} (\Sigma^\nu C^{-+})_{a_2 a_4} \nonumber \\
&&  \hskip.24truecm  +\
(\Sigma^\mu C^{-+})_{a_1 a_4} (\Sigma^\nu C^{-+})_{a_2 a_3} \Big) P_\mu P_\nu\\
\fl
\left\{Q_{a_1}^+,Q_{a_2}^+,Q_{a_3}^+,Q_{a'_4}^-\right\}&=&4i\Big(
(\Sigma^\mu C^{-+})_{a_1 a_2}C^{+-}_{a_3 a'_4} +
 (\Sigma^\mu C^{-+})_{a_1 a_3}C^{+-}_{a_2 a'_4} +
 (\Sigma^\mu C^{-+})_{a_2 a_3}C^{+-}_{a_1 a'_4}\Big) Z P_\mu \nonumber\\
\fl
\left\{Q_{a_1}^+,Q_{a_2}^+,Q_{a'_3}^-,Q_{a'_4}^-\right\}&=&\Big(
-8(\Sigma^\mu C^{-+})_{a_1 a_2} (\Sigma^\nu C^{+-})_{a'_3
a'_4}P_\mu P_\nu + C_{a_1 a_3'}^{+-} C_{a_2 a_4'}^{+-} Z^2+ C_{a_2
a_3'}^{+-} C_{a_1 a_4'}^{+-} Z^2\Big),\nonumber \eeqa plus similar
brackets involving one $Q^+$ and three $Q^-$ or four $Q^-$. We can
observe that the algebra has a very similar appearance to
(\ref{D=9-10}) (compare in particular to the second relation).

As we have mentioned previously, a representation of the
super-Poincar\'e algebra is automatically a representation of the
induced quartic algebra. Since the assumption on the non-vanishing
of the central charge seems to be more interesting (look at the
second bracket in (\ref{eq:quart}) in relation to the second
bracket of (\ref{4poin})), we only have considered the case were
the representation of supersymmetry does not trivialise the
central charge $Z$. This fact excludes the massless
representations and the massive BPS-saturated bound \cite{fsz}. In
other words, massive representations with $\left|Z\right|<2M$ show
an unexpected behaviour under quartic symmetries. But now, since
we are considering massive representations, the number of
fermionic degrees of freedom cannot exceed $N_f\le 16$. In
consequence, the only dimensions where a quartic extension leading
to non-trivial results exist are $D=2,3,4$ and $7$, respectively.
The number of supercharges are such that  the required spinors for
the extensions are like follows:
\begin{enumerate}
\item in $D=2$ one left-handed and one right-handed Majorana-Weyl
spinor; \item in $D=3,4$ two Majorana spinors; \item in $D=7$ one
$SU(2)-$Majorana spinor.
\end{enumerate}

An appealing consequence of this is that the massive invariant
$N=2$ Lagrangians constructed so far in this specific dimensions
are moreover invariant with respect to the transformations induced
by the quartic algebra. Thus, the corresponding $N=2$
supermultiplet and their associated transformations laws will
automatically be an invariant multiplet of the corresponding
quartic structure with the same transformation properties (see
footnote \P \; on page 6). Therefore, this simple observation
means that, within this construction, interacting Lagrangians
invariant under quartic symmetries are obtained. In the same sense
as the notion of supersymmetry surmounts the obstruction described
by the no-go theorem of Coleman-Mandula \cite{cm}, our present
construction can be seen as an analogous expansion that avoids the
constraints of the Haag-Lopuszanski-Sohnius theorem \cite{hls}.
This analogy should however not be misunderstood at the conceptual
level. While supersymmetry represents a fertile novelty (with
respect to classical inner/outer symmetry analysis) by means of
the introduction of fermionic charges, therefore leading to new
phenomenological aspects, the construction of quartic algebras
executed in this work is heavily dependent on the supersymmetric
algebra formalism and underlying constraints.

\subsection{Representations of quartic extensions}
The ansatz linking algebras of order four to Lie superalgebras has
remarkable consequences concerning their respective representation
theories, in the sense that superalgebra representations
automatically induce representations of the order four structures.
In full generality, the reversal of this assertion is not true, as
we briefly justify. Consider for instance the four-dimensional
quartic extensions of the Poincar\'e algebra in four space-time
dimensions. If we  study massive representations, the little
algebra is generated by the $Q$'s and $P^0=-im$ and the
four-brackets take the form \beqa \fl
\big\{Q_{\alpha_1}{}^{I_1},Q_{\alpha_2}{}^{I_2},Q_{\alpha_3}{}^{I_3},Q_{\alpha_4}{}^{I_4}\big\}&=&
2Z^2\Big(\varepsilon_{\alpha_1 \alpha_2} \varepsilon_{\alpha_3
\alpha_4} \varepsilon^{I_1 I_2} \varepsilon^{I_3 I_4}\nonumber \\
&& +\varepsilon_{\alpha_1 \alpha_3} \varepsilon_{\alpha_2 \alpha_4}
\varepsilon^{I_1 I_3} \varepsilon^{I_2 I_4} +\varepsilon_{\alpha_1
\alpha_4} \varepsilon_{\alpha_2 \alpha_3} \varepsilon^{I_1 I_4}
\varepsilon^{I_2 I_3}
\Big), \nonumber \\
\fl \left\{Q^{I_1}{}_{\alpha_1},Q^{I_2}{}_{\alpha_2},
Q^{I_3}{}_{\alpha_3}, \bar{Q}_{I_4}{}_{\dot \alpha_4} \right\}&=&
2 m Z \Big(\delta^{I_1}{}_{I_4} \varepsilon^{I_2 I_3}
\varepsilon_{\alpha_2 \alpha_3} \sigma^0{}_{\alpha_1 \dot
\alpha_4} +\delta^{I_2}{}_{I_4} \varepsilon^{I_1 I_3}
\varepsilon_{\alpha_1 \alpha_3} \sigma^0{}_{\alpha_2 \dot \alpha_4}  \nonumber \\
&+&\delta^{I_3}{}_{I_4} \varepsilon^{I_1 I_2}
\varepsilon_{\alpha_1 \alpha_2} \sigma^0{}_{\alpha_3 \dot
\alpha_4}\Big) ,
\nonumber \\
\fl \left\{Q^{I_1}{}_{\alpha_1},Q^{I_2}{}_{\alpha_2},
\bar{Q}_{I_3}{}_{\dot \alpha_3}, \bar{Q}_{I_4}{}_{\dot \alpha_4}
\right\}&=&2 m^2 \Big(\delta^{I_1}{}_{I_3}
\delta^{I_2}{}_{I_4}\sigma^0{}_{\alpha_1 \dot \alpha_3}
\sigma^0{}_{\alpha_2 \dot \alpha_4} +\delta^{I_1}{}_{I_4}
\delta^{I_2}{}_{I_3}\sigma^0{}_{\alpha_1 \dot
\alpha_4} \sigma^0{}_{\alpha_2 \dot \alpha_3} \Big) \nonumber \\
&& \hskip-.38truecm +\ 2 Z^2 \varepsilon_{\alpha_1 \alpha_2}
\varepsilon_{\dot \alpha_3 \dot \alpha_4} \varepsilon^{I_1I_2}
\varepsilon_{I_3 I_4}. \nonumber \eeqa If we now make the
following substitutions (analogous to the corresponding
substitution for the $N=2$ supersymmetric extension with central
charge): \beqa
\begin{array}{cc}
a^1=Q^1{}_1-\bar{Q}_{2 \dot 2}\ ,&a^3= Q^1{}_1+\bar{Q}_{2 \dot 2}\ , \\
a^2=Q^1{}_2+ \bar{Q}_{2 \dot 1}\ , &a^4=Q^1{}_2- \bar{Q}_{2 \dot
1} \ ,
\end{array}
\eeqa one observes that $a^1,\cdots,a^4,a^\dag_1,\cdots, a^\dag_4$
generate the Clifford algebra of the polynomial \beqa \fl
P^2(x_1,\cdots,x_4,y^1,\cdots,y^4)&=&\Big(2(2m+Z)x_1 y^1 +
2(2m+Z)x_2 y^2 +2(2m-Z)x_3 y^3 \nonumber \\ &&  + 2(2m-Z)x_4 y^4\Big)^2
\nonumber
\eeqa in the sense that \beqa \label{eq:cliff} \Big(x_I a^I + y^I
a^\dag_I\Big)^4 = P^2(x_1,\cdots,x_4,y^1,\cdots,y^4). \eeqa The
representations of the $N=2$ supersymmetric algebra in four
dimensions are obtained from the study of representations of the
Clifford algebra, {\it i.e.} when the $a$'s   satisfy the
quadratic relation, \beqa \label{eq:cliff2} \Big(x_I a^I + y^I
a^\dag_I\Big)^2 = P(x_1,\cdots,x_4,y^1,\cdots,y^4), \eeqa which is
obviously compatible with (\ref{eq:cliff}). On the contrary, one
can construct representations of (\ref{eq:cliff}) such that
(\ref{eq:cliff2}) is not satisfied. The algebra (\ref{eq:cliff})
has been introduced a long time ago by mathematicians and is
called  the Clifford algebra of the polynomial $P^2$ \cite{cp}. It
has been shown that to any polynomial $f$ one can associate a
Clifford algebra ${\cal C}_f$, and that a matrix representation
can be obtained \cite{line}. But for polynomials of degree higher
than two, the representation is not unique, and various
inequivalent representations of ${\cal C}_f$ (even of the same
dimension) can be constructed (see, for instance, \cite{ineq}).
This difficulty contributes considerably to the problem of
classifying representations of ${\cal C}_f$, which is still open,
though it has been proved that the dimension of the representation
is a multiple of the degree of the polynomial \cite{dimrep}.

In the framework of our analysis, this may provide new
representations corresponding to interesting quartic extensions of
the Poincar\'e algebra. This hierarchy of representations  on the
top of the standard representations obtained in supersymmetric
theories  might be compared to the parafermionic extension of the
Poincar\'e algebra considered in \cite{j}. It turns out that the
algebra studied there shares some similarities with ours, like the
non-linearity and the possibility of deriving a hierarchy of
representations starting from the standard supersymmetric one. To
which extent these two different approaches have additional
far-reaching common features and reflect physical phenomena in
supersymmetric theories still constitutes work in progress.

\section{Perspectives and outlook}
In this paper we have given a systematic way to associate a
quartic Lie algebra (which closes with fully symmetric quartic
brackets) to a graded Lie superalgebra of a certain type. In
particular, this construction can be applied to standard
supersymmetric theories. This specifically alludes to the fact
that any representation of $N=2$ supersymmetric algebras shares a
hidden quartic symmetry. We insist upon the fundamental role of
the non-vanishing central charge, implying that, in these
conditions, any supersymmetry Lagrangian reflects a hidden quartic
symmetry. One may wonder whether this apparently simple but
unexpected peculiarity is a consequence of some deep ``raison
d'\^{e}tre" responding to some structural or phenomenological
aspect adequately described by this type of hidden symmetry, and
somehow encoded in the conformation of the Lagrangian.

\smallskip
Another question that emerges naturally from our analysis is what
happens if we relax the assumption on the  non-vanishing of the
central charge. This opens the possibility of having massless
representations, and we are no more limited  to the few cases
exhibited in section \ref{sec:4susy}. A particularly interesting
case is that of ten-dimensional type $IIA$ supersymmetry, which
constitutes in some sense an exceptional model $-$ recall that in
this case a quartic extension is constructed with only one
Majorana spinor $-$ and inherits a hidden quartic symmetry in
quite natural way.

\smallskip

We have pointed out that the construction of the quartic algebra
structure does not impose constraints on the number of central
charges. In this sense, an analogous analysis of $N>2$
supersymmetric extensions can be considered. Also in this case,
the quartic Clifford algebras enable us to connect the underlying
representation theory of these structures. The main difficulty in
this task refers to the fact whether the $N>2$ extensions can be
written in the adequate form, in order to apply the procedure, as
well as to consider the physically relevant models. Work in this
direction is currently in progress.

\section*{Acknowledgments} During the preparation of this work, one
of the authors (RCS) was financially supported by the research
project GR58/4120818-920920 of the UCM-BSCH.

\end{document}